\documentclass{article}
\usepackage{amssymb}

\usepackage{graphicx}
\usepackage{amsmath}

\begin{document}

\title{New physics scales and anomalous magnetic moment}
\author{Helder Ch\'{a}vez$^{\text{ }\ddagger \text{ }}\thanks{%
Fellow of Centro Latinoamericano de F\'{i}sica.}$\ ,\ Luis Masperi$%
^{\maltese \text{ }}$\thanks{%
On leave of absence from Centro At\'{o}mico Bariloche, San Carlos de
Bariloche, Argentina. } \\
$^{\ddagger }$\ Centro Brasileiro de Pesquisas F\'{i}sicas,\ \ \\
Dr. Xavier Sigaud 150, 22290-180 Rio de Janeiro, Brazil\ \\
$^{\maltese }$\ Centro Latinoamericano de F\'{i}sica,\ \ \\
Av. Venceslau Br\'{a}z 71 Fundos, 22290-140 Rio de Janeiro, Brazil\ \\
}
\maketitle

\begin{abstract}
Violation\smallskip\ of chiral symmetry together with change of mass sign
\smallskip allows a linear correction in inverse power of new \smallskip
physics scale to the anomalous magnetic moment\smallskip\ of muon. In this
light we analyse alternative models \smallskip showing in particular that
grand unification, supersymmetry \smallskip and muon substructure may
explain the discrepancy \smallskip between the experimental value of the
muon anomaly \smallskip and the standard model calculation.
\end{abstract}

\section{Introduction}

\noindent The recen\thinspace \smallskip t BNL measurement of the anomalous
magnetic moment of the muon $a_{\mu }=\frac{g-2}{2}$\ has confirmed an
excess on the calculation of standard model contributions of
\begin{equation}
{\Large \Delta a}_{\mu }{\Large \simeq 2\times 10}^{-9}  \tag{1}
\end{equation}
at a $2\sigma $\ level \cite{1}.

This is an indication of possible new\smallskip\ physics at an energy scale $%
\Lambda .$\ It is interesting to estimate the order of the correction of $%
a_{\mu }$\ in \smallskip powers of $\frac{m_{\mu }}{\Lambda }$\ , where $%
m_{\mu }$\ is the muon mass. This is related \cite{2} to the validity or
breaking \smallskip of the chiral symmetry of leptons together with
the\smallskip\ change of sign of $m_{\mu }$. If this symmetry,\smallskip\
which we will call of Weinberg, is respected $\Delta a_{\mu }\backsim \left(
m_{\mu }/\Lambda \right) ^{2}$\ whereas if it is broken $\Delta a_{\mu
}\backsim m_{\mu }/\Lambda $\ . \smallskip This is important because in the
latter case the explanation of Eq. (1) may be given by \smallskip new
physics at a relatively high energy whereas in the former it should appear
at a scale close to the electroweak one.

\section{Weinberg symmetry and Standard Model}

\bigskip Since the mass term breaks chiral symmetry, one can make it
invariant changing the mass sign
\begin{equation}
{\Large \mu \rightarrow \gamma }_{5}{\Large \mu \quad ,\quad m}_{\mu }%
{\Large \rightarrow -m}_{\mu }{\Large \,.}  \tag{2}
\end{equation}
The ef\smallskip fective interaction for the anomalous magnetic moment has a
similar chiral form
\begin{equation}
\mathcal{L}_{\mathbf{eff}}\ {\Large =a}_{\mu }{\Large \,}\frac{e}{4m_{\mu }}%
\overline{\mu }{\Large \sigma }^{\alpha \beta }{\Large \mu F}_{\alpha \beta }%
{\Large \quad ,}  \tag{3}
\end{equation}
so that if the symmetry Eq. \smallskip (2) is valid the corrections to $%
a_{\mu }$\ must be of even powers of the ratio of $m_{\mu }$\ to a larger
scale $\Lambda $%
\begin{equation}
{\Large a}_{\mu }{\Large =c}_{o}\left( \frac{m_{\mu }}{\Lambda }\right) ^{0}%
{\Large +c}_{2}\left( \frac{m_{\mu }}{\Lambda }\right) ^{2}{\Large +...\quad
.}  \tag{4}
\end{equation}

This is\smallskip\ what happens in the Standard Model (SM) since, to be
invariant in it, Eq. (3) must be written as
\begin{equation}
\mathcal{L}_{\mathbf{eff}}\ {\Large =a}_{\mu }{\Large \,}\frac{e}{4m_{\mu }}%
\left( \overline{l}_{L}\sigma ^{\alpha \beta }\mu _{R}\frac{f\varphi _{V}}{%
m_{\mu }}+h.c.\right) {\Large F}_{\alpha \beta }{\Large \quad ,}  \tag{5}
\end{equation}
with a doublet Higgs field $\varphi =\varphi _{V}+\left(
\begin{array}{c}
0 \\
h/\sqrt{2}
\end{array}
\right) $ such that
\begin{equation}
\overline{l}_{L}\ {\Large =}\left(
\begin{array}{c}
\nu _{L} \\
\mu _{L}
\end{array}
\right) {\Large \quad ,\quad \varphi }_{V}\ {\Large =}\left(
\begin{array}{c}
0 \\
\frac{\upsilon }{\sqrt{2}}
\end{array}
\right) {\Large \quad ,\,\,}f\frac{\upsilon }{\sqrt{2}}\ {\Large =\ }m_{\mu
}\,{\Large \,.}  \tag{6}
\end{equation}
To have the transformations Eq. (2) one must perform
\begin{equation}
l_{L}{\Large \rightarrow \gamma }_{5}{\Large \,}l_{L}{\Large =-}l_{L}{\Large %
\quad ,\quad \mu }_{R}{\Large \rightarrow \gamma }_{5}{\Large \mu }_{R}%
{\Large =\mu }_{R}{\Large \quad ,\quad \varphi \rightarrow -\varphi \ .}
\tag{7}
\end{equation}

The W\smallskip einberg symmetry (WS) is respected in SM since the charged
weak interactions
\begin{equation*}
\overline{\nu }_{L}{\Large \gamma }^{\alpha }{\Large \mu }_{L}{\Large W}%
_{\alpha }{\Large \ ,}
\end{equation*}
neutral and electromagnetic
\begin{equation*}
\overline{\mu }{\Large \gamma }^{\alpha }\left( g_{\upsilon }+g_{a}\gamma
^{5}\right) {\Large \mu Z}_{\alpha }{\Large \quad ,\quad }\overline{\mu }%
{\Large \gamma }^{\alpha }{\Large \mu A}_{\alpha }
\end{equation*}
and Yukawa one
\begin{equation*}
\overline{\mu }{\Large \mu \,h\ ,}
\end{equation*}
are invariant \smallskip under Eq. (7) .\newline
Therefore the \smallskip corrections to $a_{\mu }$\ are of the type of Eq.
(4) with the electroweak scale $\Lambda _{EW}$\ , corresponding \smallskip
the first term to the electromagnetic contributions $c_{o}=\frac{\alpha }{%
2\pi }+...$\ and the second to the weak ones $\backsim 10^{-9}$\ .

\section{Grand Unification Theories}

In principle, if there is\smallskip\ a second \cite{3} Higgs doublet $%
\varphi
{\acute{}}%
$\ wich breaks at a scale $\Lambda $\ higher\smallskip\ than that of SM and
is not \smallskip directly related to the muon mass, it is possible to have
an additional effective interaction of new physics
\begin{equation}
\mathcal{L}_{\mathbf{eff}}\ {\Large \backsim \ }\frac{1}{\Lambda ^{2}}\left(
\overline{l}_{L}\sigma ^{\alpha \beta }\mu _{R}\varphi
{\acute{}}%
_{V}+h.c.\right) {\Large F}_{\alpha \beta }{\Large \quad .}  \tag{8}
\end{equation}
Being $\varphi
{\acute{}}%
_{V}\backsim \Lambda $\ , Eq. (8) is of the form
\begin{equation}
\mathcal{L}_{\mathbf{eff}}\ {\Large =\Delta a}_{\mu }\frac{e}{4m_{\mu }}%
\overline{\mu }{\Large \sigma }^{\alpha \beta }{\Large \mu F}_{\alpha \beta }
\tag{9}
\end{equation}
with $\Delta a_{\mu }=O\left( \frac{m_{\mu }}{\Lambda }\right) $\ \smallskip
which breaks the\smallskip\ Weinberg symmetry.

Regarding Grand\smallskip\ \smallskip Unification Theories(GUT) it is not
obvious that a particular model has interactions \smallskip which\smallskip\
violate this symmetry and therefore produce an effective interaction like
Eq. (8).\smallskip

E.g. \smallskip starting with SU(5), the only new interaction which
contributes to $a_{\mu }$\ is that due to leptoquarks $X^{\alpha }$\ i.e. $%
\overline{\mu }\gamma ^{\alpha }qX_{\alpha }$\ which\smallskip\ \smallskip
changes muon into quarks but respects chiral symmetry $\mu \rightarrow
\gamma _{5}\mu $\ , $q\rightarrow \gamma _{5}q.$\ Since the Yukawa\smallskip
\smallskip\ interaction is unchanged because the Higgs\smallskip\ which
breaks $SU(5)$\ does not affect the\smallskip\ muon mass, the WS is still
valid, the effective interaction is of the type of Eq. (5) and
\begin{equation}
{\Large \Delta a}_{\mu }^{SU(5)}\ {\Large \backsim \,}\left( \frac{m_{\mu }}{%
\Lambda }\right) ^{2}{\Large ,}  \tag{10}
\end{equation}
which is negligible for the\smallskip\ GUT \smallskip scale $\backsim
10^{15}GeV.$

In the case of $SO(10),$\ the only\smallskip\ change is that \smallskip the
high-scale Higgs $\varphi
{\acute{}}%
$\ gives mass to $\nu _{R}$\ with a Majorana term.\smallskip\ Then the low
scale\smallskip\ Higgs $\varphi $\ gives also to $\nu $\ a Dirac mass. The
diagonalization\smallskip\ of the mass matrix \smallskip produces the
see-saw mechanism which involves\smallskip\ a mixture of $\nu _{L}$\ and $%
\nu _{R}$. But this only means \smallskip that also the charged weak
interactions \smallskip have a small right-lepton contribution. Since $%
\overline{\nu }\gamma ^{\alpha }\mu W_{\alpha }$\ is\smallskip\ chiral
invariant and the Yukawa muon term $\overline{\mu }\mu h$\ is unchanged, the
WS is preserved and again
\begin{equation*}
{\Large \Delta a}_{\mu }^{SO(10)}\ {\Large \backsim \ }\left( \frac{m_{\mu }%
}{\Lambda }\right) ^{2}{\Large .}
\end{equation*}

The substantial difference of the\smallskip\ exceptional group $E_{6}$\ is
that it has 11 additional superheavy fermions \smallskip among which
a\smallskip\ charged lepton $M$\ that can mix with $\mu .\smallskip $

If the \smallskip breakings of symmetry are\smallskip\ due to \cite{4} a 351
of $E_{6}$\ , when GUT is broken the \smallskip mass eigenstates $\mu _{o}$\
(massless) \smallskip and $\widehat{M}$\ (superheavy) are determined by the
expectation\smallskip\ values of the $\left( SO(10),SU(5)\right) $\
multiplets\smallskip\ $\varphi \,%
{\acute{}}%
\,\left( \mathbf{54,24}\right) $\ and $\varphi \,%
{\acute{}}%
\,\left( \mathbf{144,24}\right) $\ through a large mixture of left
components
\begin{equation}
{\Large \mu }_{L}^{o}{\Large =\mu }_{L}\cos {\Large \theta }_{L}{\Large +M}%
_{L}\sin {\Large \theta }_{L}{\Large \quad ,\quad }\widehat{M}_{L}{\Large =M}%
_{L}\cos {\Large \theta }_{L}{\Large -\mu }_{L}\sin {\Large \theta }_{L}
\tag{12}
\end{equation}
with the same mixing of\smallskip\ $\nu _{L}$\ and the neutral exotic lepton
$N_{L}.$

The small mass of\smallskip\ ordinary muon is due to the appearance of an
expectation value of a Higgs $H(10,\overline{\mathbf{5}})$\ which\smallskip\
in terms of the doublet $\varphi $\ gives
\begin{equation}
\overline{l}_{L}{\Large \mu }_{R}{\Large f\,\varphi +h.c.=}\frac{f}{\sqrt{2}}%
\left( \upsilon +h\right) \left( \overline{\mu }_{R}\mu _{L}+h.c.\right)
{\Large \,.}  \tag{13}
\end{equation}

Since the\smallskip\ right components were not mixed at the GUT scale, the
mass term from Eq. (13) is
\begin{equation*}
\frac{f}{\sqrt{2}}{\Large \upsilon }\left( \overline{\mu }_{R}^{o}\mu
_{L}^{o}\cos \theta _{L}-\overline{\mu }_{R}^{o}\widehat{M}_{L}\sin \theta
_{L}+h.c.\right) {\Large \,.}
\end{equation*}
Diagonalizing the whole \smallskip mass matrix to give the physical state $%
\widehat{\mu }$\ there will be also a mixture of $\mu _{R}^{o}$\ with $%
\widehat{M}_{R}$\ \smallskip which will\smallskip\ be small due to the very
different GUT and EW scales, but with\smallskip\ a relevant contribution
to\smallskip\ the muon mass. Therefore it will not be possible to argue that
the transformation $\varphi \rightarrow -\varphi $\ will assure
the\smallskip\ change\smallskip\ $m_{\mu }\rightarrow -m_{\mu }$\ .

Regarding the interaction\smallskip\ with the\smallskip\ light Higgs $h$\ ,
there will be a ``flavour-changing'' term approximately equal to
\begin{equation}
\mathcal{L}_{\mathbf{eff}}\ {\Large =-}\frac{f\,\,h}{\sqrt{2}}\left(
\overline{\widehat{\mu }}_{R}\widehat{M}_{L}\sin \theta _{L}+h.c.\right)
{\Large \,,}  \tag{14}
\end{equation}
due to the slight difference \smallskip between $\mu ^{o}$\ and the physical
state $\widehat{\mu }\,.$\ \smallskip This interaction does not respect WS
because even though it is invariant under
\begin{equation}
\widehat{\mu }{\Large \rightarrow \gamma }_{5}\widehat{\mu }{\Large \quad
,\quad }\widehat{M}{\Large \rightarrow \gamma }_{5}\widehat{M}{\Large \quad
,\quad h\rightarrow -h\quad ,}  \tag{15}
\end{equation}
as said above the last\smallskip\ transformation does not imply $m_{\mu
}\rightarrow -m_{\mu }$\ . As a\smallskip\ consequence one may expect a
linear correction\smallskip\ of the muon magnetic moment. \smallskip In fact
the explicit calculation \cite{5} of the one-loop contribution caused by Eq.
(14) gives
\begin{equation}
{\Large \Delta a}_{\mu }^{FCh}\ {\Large \simeq \ }\frac{1}{16\pi ^{2}}%
{\Large \kappa }^{2}\frac{m_{\mu }}{M_{M}}{\Large \quad ,}  \tag{16}
\end{equation}
where, being $\kappa =\frac{1}{\sqrt{2}}f\,\sin \theta _{L}\lesssim 1$\ and
here $f$\ $\,$not necessarily \smallskip as small as $\sqrt{2}\frac{m_{\mu }%
}{\upsilon }$, to explain the discrepancy Eq. (1) \smallskip one needs $%
M_{M}\lesssim 10^{6}$\ $GeV.$\ \smallskip Even though this mass value seems
small for a GUT\smallskip\ particle, it is not unreasonable considering
\smallskip the strong mixture of exotic and\smallskip\ ordinary fermions.
The interaction Eq.(14) we have deduced \smallskip from a particular
\smallskip scheme of mixture of muon with exotic lepton of $E_{6}$\ is
equivalent to the one\smallskip\ of a singlet charged heavy lepton with the
muonic and light Higgs \smallskip doublets \cite{6}.

The mixture of $\mu $\ and $M$\ \smallskip produces additional corrections
to $a_{\mu }$\ \smallskip of electroweak type . If the scheme of breakings
is\smallskip\ based on the multiplet 35\smallskip 1 as said above, the equal
mixings for $\mu _{L}$\ and $\nu _{L}$\ \smallskip avoid any
correction\smallskip\ in the charged current interaction. The same
happens\smallskip\ for the neutral charge interaction with\smallskip\ $Z$\
if only the left mixtures are considered.\smallskip\ But if the small right
mixture is included, a coupling $\overline{\widehat{\mu }}\gamma ^{\alpha }%
\widehat{M}Z_{\alpha }$\ appears which is however chiral invariant and gives
\begin{equation}
{\Large \Delta a}_{\mu }^{FCZ}\ {\Large \simeq \ }\frac{\alpha }{\pi }\left(
\sin \theta _{R}\cos \theta _{R}\right) ^{2}\left( \frac{m_{\mu }}{M_{Z}}%
\right) ^{2}{\Large ,}  \tag{17}
\end{equation}
where $\left( \sin \theta _{R}\cos \theta _{R}\right) ^{2}<10^{-2}$\
\smallskip not to spoil the experimental $\mu \mu Z$\ coupling \cite{7}, so
that $\Delta a_{\mu }^{FCZ}\lesssim 10^{-11}$\ i.e.\smallskip\ irrelevant.

For the \smallskip correction \smallskip given by a loop with exchange of a
heavy neutral boson $Z%
{\acute{}}%
,$\ with different\smallskip\ charges\smallskip\ of the corresponding
abelian group for $\mu $\ and $M$\ , there will be a ``flavour changing''
contribution
\begin{equation}
{\Large \Delta a}_{\mu }^{FCZ\,%
{\acute{}}%
}{\Large \simeq }\frac{\alpha }{\pi }\left( \sin \theta _{L}\cos \theta
_{L}\right) ^{2}\left( \frac{m_{\mu }}{M_{Z\,%
{\acute{}}%
}}\right) ^{2}{\Large +}\frac{\alpha }{\pi }\sin {\Large \theta }_{L}\cos
{\Large \theta }_{L}\sin {\Large \theta }_{R}\cos {\Large \theta }_{R}\frac{%
m_{\mu }}{M_{Z%
{\acute{}}%
}}  \tag{18}
\end{equation}
where the first chiral \smallskip conserving term depends only on the large
left mixture and is quadratic in the new physics\smallskip\ scale, whereas
the second chiral interference term needs the small right mixture
reflecting\smallskip\ the two-scales building of mass states and\smallskip\
is linear in $m_{\mu }/M_{Z\,%
{\acute{}}%
}$\ .

But, compared\smallskip\ with Eq. (16),\smallskip\ even if $M_{Z\,\,%
{\acute{}}%
}$\ is so low as $10^{6}GeV$\ this linear term gives again $\Delta a_{\mu
}\lesssim 10^{-11}$\ i.e. negligible.

\section{Other theories beyond the Standard Model}

One of t\smallskip he most obvious is the Minimal Supersymmetric Standard
Model (MSSM) which has two \smallskip Higgs doublets with a ratio of
expectation values $\tan \beta =\frac{\left\langle \,H\,\,%
{\acute{}}%
\,\right\rangle }{\left\langle \,H\ \,\right\rangle }$\ where $H$\ gives
mass to electron and quark \smallskip $d$, and $H%
{\acute{}}%
$\ to quark $u$.

The interaction\smallskip\ of muon with the chargino $\widetilde{W}$\ , $%
\overline{\mu }\widetilde{W}\widetilde{\nu }$\ , in a way similar to Eq.
(18), gives \cite{8} the correction of $a_{\mu }$\
\begin{equation}
{\Large \Delta a}_{\mu }^{\widetilde{W}}{\Large \backsim }\frac{\alpha }{\pi
}\left( \frac{m_{\mu }}{M_{\widetilde{W}}}\right) ^{2}{\Large +}\frac{\alpha
}{\pi }\frac{m_{\mu }}{M_{\widetilde{W}}}{\Large O}\left( \frac{m_{\mu }\tan
\beta }{M_{W}}\right) {\Large \ ,}  \tag{19}
\end{equation}
for large $\tan \beta ,$\ \smallskip where \smallskip the first quadratic
term corresponds to conservation of chirality and the second linear one
is\smallskip\ due to the chiral violation in internal line caused by the
coupling\smallskip\ of the two Higgs \cite{9}.

An analogous correction $\Delta a_{\mu }^{\widetilde{Z}}$\ comes from the
coupling $\overline{\mu }\widetilde{Z}\widetilde{\mu }$\ re\smallskip
placing in Eq. (19) $M_{\widetilde{W}}$\ by $M_{\widetilde{Z}}.$\ Since
\smallskip one\smallskip\ expects $M_{\widetilde{W}}$\ $\backsim M_{%
\widetilde{Z}}\backsim 1$\ $TeV$\ , it is possible that these two chiral
violating contributions \smallskip add to $\Delta a_{\mu }^{SUSY}\backsim
10^{-9}.$

The hypothesis\smallskip \smallskip\ of a large extra dimension, and a
simple Higgs doublet, gives way to a strong\smallskip\ gravitation exchange
producing for a relevant number of Kaluza-Klein states \cite{10}
\begin{equation}
\mathcal{L}_{\mathbf{eff}}^{ED}\ {\Large \backsim \,}\left( \frac{m_{\mu }}{%
\Lambda }\right) ^{2}\frac{e}{4m_{\mu }}\overline{\mu }{\Large \sigma }%
^{\alpha \beta }{\Large \mu F}_{\alpha \beta }{\Large \ ,}  \tag{20}
\end{equation}
that to satisfy the\smallskip\ discrepancy Eq. (1) would require $\Lambda
\backsim TeV$\ which seems excluded by\smallskip\ astrophysical\smallskip\
observations. An analogous contribution may come from the exchange of an
antisymmetric tensor\smallskip\ field encountered in string theory \cite{11}.

It is interesting that a model \cite{12} of breaking of SUSY through
boundary conditions in a 5th dimension of \smallskip radius $R\backsim
TeV^{-1}$ allows to use only one Higgs doublet which would modify the
contribution to $a_{\mu }$ given by \smallskip Eq. (19).

The possibility \smallskip of a new abelian gauge symmetry felt by muon and
tauon but not by electron \cite{13} would give an additional quadratic
contribution to $a_{\mu }$\ similar to that of the exchange of $Z$\ of the
\smallskip second term of Eq. (4) which might be $\backsim 10^{-9}$. Mixing
of the two neutral gauge bosons may \smallskip produce \cite{14} a linear
correction of $a_{\mu }.$

Finally, the \smallskip models\smallskip\ of substructure of muons are
different in the sense that they do not \smallskip correspond simply to add
new particles to those of the SM. Their effect to $a_{\mu }$\ depends on the
model \smallskip assumptions.

One\smallskip\ reasoning \cite{15} is that since the terms of mass and
anomalous magnetic moment of \smallskip muon have the same chiral
structure,\smallskip\ the scale of substructure should enter\smallskip\ into
them as $\Lambda \overline{\mu }\mu $\ and $\frac{e}{\Lambda }\overline{\mu }%
\sigma ^{\alpha \beta }\mu F_{\alpha \beta }$\ , respectively.
One\smallskip\ would expect that in the\smallskip\ same way as a decreasing
factor $\frac{m_{\mu }}{\Lambda }$\ perhaps related to a chiral\smallskip
\smallskip\ symmetry must be introduced in the former, a similar one in the
latter should give\smallskip\ $\Delta a_{\mu }\backsim \left( \frac{m_{\mu }%
}{\Lambda }\right) ^{2}.$

However, \smallskip a \smallskip specific model \cite{16} which assumes the
appearance at a scale $\Lambda $\ of a \smallskip Yukawa-type \smallskip
vertex $\mu ^{-}\rightarrow \pi ^{-}N$\ , with a heavy neutrino $N,$\ gives
a linear \smallskip correction $\Delta a_{\mu }\backsim \frac{m_{\mu }}{%
\Lambda }$\ since the WS is violated because the pion field has nothing to
do with $m_{\mu }.$

\section{Conclusion}

We have\smallskip\ seen that interactions which break chiral symmetry
together with change of sign of muon mass \smallskip favour the contribution
of\smallskip\ new physics to the anomalous magnetic moment $a_{\mu }.$\ In
this\smallskip\ way,\smallskip\ parameter regions of MSSM, additional
neutral gauge interactions, \smallskip particular models of muon\smallskip\
substructure, or even a GUT alternative of strong mixing of
ordinary\smallskip\ left lepton doublet \smallskip with the exotic one of $%
E_{6}$\ \smallskip together with a small mixture of the corresponding right
charged\smallskip\ component, may\smallskip\ explain the discrepancy between
the recent experimental measurement \smallskip of $a_{\mu }$\ and its
calculation within Stan\smallskip \smallskip dard Model. It is obviously too
soon to say that the need\smallskip\ of\smallskip\ physics beyond the SM has
been proved \smallskip since the above discrepancy is so far only a 2$\sigma
$\ effect.\smallskip \qquad

\begin{quotation}
\smallskip \textbf{Acknowledgements}
\end{quotation}

\smallskip The authors thank Ernest Ma for interesting discussions. The work
of H. Ch. was financially supported by CNPq of Brazil.

\end{document}